# Five Misunderstandings About Case-Study Research

Bent Flyvbjerg

**Abstract**

This article examines five common misunderstandings about case-study research: (1) Theoretical knowledge is more valuable than practical knowledge; (2) One cannot generalize from a single case, therefore the single case study cannot contribute to scientific development; (3) The case study is most useful for generating hypotheses, while other methods are more suitable for hypotheses testing and theory building; (4) The case study contains a bias toward verification; and (5) It is often difficult to summarize specific case studies. The article explains and corrects these misunderstandings one by one and concludes with the Kuhnian insight that a scientific discipline without a large number of thoroughly executed case studies is a discipline without systematic production of exemplars, and that a discipline without exemplars is an ineffective one. Social science may be strengthened by the execution of more good case studies.

**Introduction**

When I first became interested in in-depth case-study research, I was trying to understand how power and rationality shape each other and form the urban environments in which we live (Flyvbjerg 1998a). It was clear to me that in order to understand a complex issue like this, in-depth case-study research was necessary. It was equally clear, however, that my teachers and colleagues kept dissuading me from employing this particular research methodology.

'You cannot generalize from a single case,' some would say, 'and social science is about generalizing.' Others would argue that the case study may be well suited for pilot studies but not for full-fledged research schemes. Others again would comment that the case study is subjective, giving too much scope for the researcher's own interpretations. Thus the validity of case studies would be wanting, they argued.

At first, I did not know how to respond to such claims, which clearly formed the conventional wisdom about case-study research. I decided therefore to find out where the claims come from and whether they are correct. This article contains what I discovered.

**The Conventional Wisdom About Case-study research**

Looking up 'case study' in the *Dictionary of Sociology* as a beginning, I found the following in full citation:

> *Case Study*. The detailed examination of a single example of a class of phenomena, a case study cannot provide reliable information about the broader class, but it may be useful in the preliminary stages of an investigation since it provides hypotheses, which may be tested systematically with a larger number of cases (Abercrombie et al. 1984, 34).[i]

This description is indicative of the conventional wisdom of case-study research, which, if not directly wrong, is so oversimplified as to be grossly misleading. It is correct that the case study is a 'detailed examination of a single example,' but as we will see below it is not true that a case study 'cannot provide reliable information about the broader class.' It is also correct that a case study *can* be used 'in the preliminary stages of an investigation' to generate hypotheses, but it is misleading to see the case study as a pilot method to be used only in preparing the real study's larger surveys, systematic hypotheses testing, and theory building.

According to the conventional view, a case and a case study cannot be of value in and of themselves; they need to be linked to hypotheses, following the well-known hypothetico-deductive model of explanation. Mattei Dogan and Dominique Pelassy (1990, 121) put it like this: 'one can validly explain a particular case only on the basis of general hypotheses. All the rest is uncontrollable, and so of no use' (see also Diamond 1996, 6). Similarly, the early Donald Campbell did not mince words when he relegated single-case studies to the methodological trash heap:

> [S]uch studies have such a total absence of control as to be of almost no scientific value . . . Any appearance of absolute knowledge, or intrinsic knowledge about singular isolated objects, is found to be illusory upon analysis . . . It seems well-nigh unethical at the present time to allow, as theses or dissertations in education, case studies of this nature (i.e., involving a single group observed at one time only) (Campbell and Stanley 1966, 6-7).

If you read such criticism of a certain methodology enough times, or if you hear your thesis advisers repeat it, you begin to believe it may be true. This is what happened to me, and it made me uncertain about case study methodology. As I continued my research, however, I found out that Campbell had later made a 180-degree turn in his views of the case study and had become one of the strongest proponents of this method. I eventually found, with the help of Campbell's later works and other works like them, that the problems with the conventional wisdom about case-study research can be summarized in five misunderstandings or oversimplifications about the nature of such research:

> Misunderstanding no. 1. General, theoretical (context-independent) knowledge is more valuable than concrete, practical (context-dependent) knowledge.

> Misunderstanding no. 2. One cannot generalize on the basis of an individual case; therefore, the case study cannot contribute to scientific development.

> Misunderstanding no. 3. The case study is most useful for generating hypotheses; that is, in the first stage of a total research process, while other methods are more suitable for hypotheses testing and theory building.
>
> Misunderstanding no. 4. The case study contains a bias toward verification, that is, a tendency to confirm the researcher's preconceived notions.
>
> Misunderstanding no. 5. It is often difficult to summarize and develop general propositions and theories on the basis of specific case studies.

These five misunderstandings indicate that it is theory, reliability, and validity which are at issue; in other words, the very status of the case study as a scientific method. In what follows, I will focus on these five misunderstandings and correct them one by one. First, however, I will outline the role of cases in human learning.

**The Role of Cases in Human Learning**

In order to understand why the conventional view of case-study research is problematic, we need to grasp the role of cases and theory in human learning. Here two points can be made. First, the case study produces the type of context-dependent knowledge which research on learning shows to be necessary to allow people to develop from rule-based beginners to virtuoso experts. Second, in the study of human affairs, there appears to exist only context-dependent knowledge, which thus presently rules out the possibility of epistemic theoretical construction. The full argument behind these two points can be found in Flyvbjerg (2001, chapters 2-4). For reasons of space, I can only give an outline of the argument here. At the outset, however, we can assert that if the two points are correct, it will have radical consequences for the conventional view of the case study in research and teaching. This view would then be problematic.

Phenomenological studies of human learning indicate that for adults there exists a qualitative leap in their learning process from the rule-governed use of analytical rationality in beginners to the fluid performance of tacit skills in what Pierre Bourdieu (1977) calls virtuosos and Hubert and Stuart Dreyfus (1986) true human experts. Here we may note that most people are experts in a number of everyday social, technical, and intellectual skills like giving a gift, riding a bicycle, or interpreting images on a television screen, while only few reach the level of true expertise for more specialized skills like playing chess, composing a symphony, or flying a fighter jet.

Common to all experts, however, is that they operate on the basis of intimate knowledge of several thousand concrete cases in their areas of expertise. Context-dependent knowledge and experience are at the very heart of expert activity. Such knowledge and expertise also lie at the center of the case study as a research and teaching method; or to put it more generally, still: as a method of learning. Phenomenological studies of the learning process therefore emphasize the importance of this and similar methods: it is only because of experience with cases that one can at all move from being a beginner to being an expert. If people were exclusively trained in context-independent knowledge and rules, that is, the kind of knowledge which forms the basis of textbooks and computers they would remain at the beginner's level in the learning process. This is the limitation of analytical rationality: it is inadequate for the best results in the exercise of a profession, as student, researcher, or practitioner.

In a teaching situation, well chosen case studies can help the student achieve competence, while context-independent facts and rules will bring the student just to the beginner's level. Only few institutions of higher learning have taken the consequence of this. Harvard University is one of them. Here both teaching and research in the professional schools are modeled to a wide extent on the understanding that case knowledge is central to human learning (Christensen and Hansen eds. 1987; Cragg 1940).

At one stage in my research, I was invited to Harvard to learn about case methodology 'in action.' During my stay, it became clear to me that if I was going to aspire at becoming an expert in my field of expertise, and if I wanted to be an effective help to my students in their learning processes, I would need to master case methodology in research and teaching. My stay at Harvard also became a major step forward in shedding my uncertainties about the conventional wisdom about cases and case studies. At Harvard I found the literature and people who effectively argued, 'Forget the conventional wisdom, go ahead and do a case study.' I figured if it is good enough for Harvard, it is good enough for me, and I suggest others might reason like this, including whole institutions of learning. There is much to gain, for instance, by transforming the lecture format still dominant in most universities to one of case learning (Christensen and Hansen eds. 1987).

It is not that rule-based knowledge should be discounted: it is important in every area and especially to novices. But to make rule-based knowledge the highest goal of learning is regressive. There is a need for both approaches. The highest levels in the learning process, that is, virtuosity and true expertise, are reached only via a person's own experiences as practitioner of the relevant skills. Therefore, beyond using the case method and other experiential methods for teaching, the best that teachers can do for students in professional programs is to help them achieve real practical experience; for example, via placement arrangements, internships, summer jobs, and the like.

For researchers, the closeness of the case study to real-life situations and its multiple wealth of details are important in two respects. First, it is important for the development of a nuanced view of reality, including the view that human behavior cannot be meaningfully understood as simply the rule-governed acts found at the lowest levels of the learning process, and in much theory. Second, cases are important for researchers' own learning processes in developing the skills needed to do good research. If researchers wish to develop their own skills to a high level, then concrete, context-dependent experience is just as central for them as to professionals learning any other specific skills. Concrete experiences can be achieved via continued proximity to the studied reality and via feedback from those under study. Great distance to the object of study and lack of feedback easily lead to a stultified learning process, which in research can lead to ritual academic blind alleys, where the effect and usefulness of research becomes unclear and untested. As a research method, the case study can be an effective remedy against this tendency.

The second main point in connection with the learning process is that there does not and probably cannot exist predictive theory in social science. Social science has not succeeded in producing general, context-independent theory and has thus in the final instance nothing else to offer than concrete, context-dependent knowledge. And the case study is especially well suited to produce this knowledge. In his later work, Donald Campbell (1975, 179) arrives at a similar conclusion, explaining how his work has undergone 'an extreme oscillation away from my earlier dogmatic disparagement of case studies,' which was described above. In a logic that in many ways resembles that of the phenomenology of human learning, Campbell now explains:

> After all, man is, in his ordinary way, a very competent knower, and qualitative common-sense knowing is not replaced by quantitative knowing . . . This is not to say that such common sense naturalistic observation is objective, dependable, or unbiased. But it is all that we have. It is the only route to knowledge--noisy, fallible, and biased though it be (1975, 179, 191).

Campbell is not the only example of a researcher who has altered his views about the value of the case study. Hans Eysenck (1976, 9), who originally did not regard the case study as anything other than a method of producing anecdotes, later realized that 'sometimes we simply have to keep our eyes open and look carefully at individual cases--not in the hope of proving anything, but rather in the hope of learning something!' Proof is hard to come by in social science because of the absence of 'hard' theory, whereas learning is certainly possibly. More recently, similar views have been expressed by Charles Ragin, Howard Becker, and their colleagues in explorations of what the case study is and can be in social inquiry (Ragin and Becker 1992).

As for predictive theory, universals, and scientism, the study of human affairs is thus at an eternal beginning. In essence, we have only specific cases and context-dependent knowledge. The first of the five misunderstandings about the case study--that general theoretical (context-independent) knowledge is more valuable than concrete, practical (context-dependent) knowledge--can therefore be revised as follows:

> *Predictive theories and universals cannot be found in the study of human affairs. Concrete, context-dependent knowledge is therefore more valuable than the vain search for predictive theories and universals.*

**Cases as 'Black Swans'**

The view that one cannot generalize on the basis of a single case is usually considered to be devastating to the case study as a scientific method. This second misunderstanding about the case study is typical among proponents of the natural science ideal within the social sciences. Yet even researchers who are not normally associated with this ideal may be found to have this viewpoint. According to Anthony Giddens, for example,

> Research which is geared primarily to hermeneutic problems may be of generalized importance in so far as it serves to elucidate the nature of agents' knowledgeability, and thereby their reasons for action, across a wide range of action-contexts. Pieces of ethnographic research like . . . say, the traditional small-scale community research of fieldwork anthropology--are not in themselves generalizing studies. But they can easily become so if carried out in some numbers, so that judgements of their typicality can justifiably be made (1984, 328).

It is correct that one can generalize in the ways Giddens describes, and that often this is both appropriate and valuable. But it would be incorrect to assert that this is the only way to work, just as it is incorrect to conclude that one cannot generalize from a single case. It depends upon the case one is speaking of, and how it is chosen. This applies to the natural sciences as well as to the study of human affairs (see also Platt 1992; Ragin and Becker 1992).

For example, Galileo's rejection of Aristotle's law of gravity was not based upon observations 'across a wide range,' and the observations were not 'carried out in some numbers.' The rejection consisted primarily of a conceptual experiment and later on of a practical one. These experiments, with the benefit of hindsight, are self-evident. Nevertheless, Aristotle's view of gravity dominated scientific inquiry for nearly two thousand years before it was falsified. In his experimental thinking, Galileo reasoned as follows: if two objects with the same weight are released from the same height at the same time, they will hit the ground simultaneously, having fallen at the same speed. If the two objects are then stuck together into one, this object will have double the weight and will according to the Aristotelian view therefore fall faster than the two individual objects. This conclusion operated in a counter-intuitive way for Galileo. The only way to avoid the contradiction was to eliminate weight as a determinant factor for acceleration in free fall. And that was what Galileo did. Historians of science continue to discuss whether Galileo actually conducted the famous experiment from the leaning tower of Pisa, or whether it is simply a myth. In any event, Galileo's experimentalism did not involve a large random sample of trials of objects falling from a wide range of randomly selected heights under varying wind conditions, and so on, as would be demanded by the thinking of the early Campbell and Giddens. Rather, it was a matter of a single experiment, that is, a case study, if any experiment was conducted at all. (On the relation between case studies, experiments, and generalization, see Lee 1989; Wilson 1987; Bailey 1992; Griffin et al. 1991). Galileo's view continued to be subjected to doubt, however, and the Aristotelian view was not finally rejected until half a century later, with the invention of the air pump. The air pump made it possible to conduct the ultimate experiment, known by every pupil, whereby a coin or a piece of lead inside a vacuum tube falls with the same speed as a feather. After this experiment, Aristotle's view could be maintained no longer. What is especially worth noting in our discussion, however, is that the matter was settled by an individual case due to the clever choice of the extremes of metal and feather. One might call it a critical case: for if Galileo's thesis held for these materials, it could be expected to be valid for all or a large range of materials. Random and large samples were at no time part of the picture. Most creative scientists simply do not work this way with this type of problem.

Carefully chosen experiments, cases, and experience were also critical to the development of the physics of Newton, Einstein, and Bohr, just as the case study occupied a central place in the works of Darwin, Marx, and Freud. In social science, too, the strategic choice of case may greatly add to the generalizability of a case study. In their classical study of the 'affluent worker,' John Goldthorpe et al. (1968, 1969) deliberately looked for a case that was as favorable as possible to the thesis that the working class, having reached middle-class status, was dissolving into a society without class identity and related conflict (see also Wieviorka 1992). If the thesis could be proved false in the favorable case, then it would most likely be false for intermediate cases. Luton, a prosperous industrial center with companies known for high wages and social stability--fertile ground for middle-class identity-- was selected as a case, and through intensive fieldwork the researchers discovered that even here an autonomous working-class culture prevailed, lending general credence to the thesis of the persistence of class identity. Below we will discuss more systematically this type of strategic sampling.

As regards the relationship between case studies, large samples, and discoveries, W. I. B. Beveridge (1951; here quoted from Kuper and Kuper eds. 1985, 95) observed immediately prior to the breakthrough of the quantitative revolution in the social sciences: '[M]ore discoveries have arisen from intense observation than from statistics applied to large groups.' This does not mean that the case study is always appropriate or relevant as a research method, or that large random samples are without value (see also the Conclusions below). The choice of method should clearly depend on the problem under study and its circumstances.

Finally, it should be mentioned that formal generalization, be it on the basis of large samples or single cases, is considerably overrated as the main source of scientific progress. Economist Mark Blaug (1980)--a self-declared adherent to the hypothetico-deductive model of science--has demonstrated that while economists typically pay lip service to the hypothetico-deductive model and to generalization, they rarely practice what they preach in actual research. More generally, Thomas Kuhn has shown that the most important precondition for science is that researchers possess a wide range of practical skills for carrying out scientific work. Generalization is just one of these. In Germanic languages, the term 'science' (*Wissenschaft*) means literally 'to gain knowledge.' And formal generalization is only one of many ways by which people gain and accumulate knowledge. That knowledge cannot be formally generalized does not mean that it cannot enter into the collective process of knowledge accumulation in a given field or in a society. A purely descriptive, phenomenological case study without any attempt to generalize can certainly be of value in this process and has often helped cut a path toward scientific innovation. This is not to criticize attempts at formal generalization, for such attempts are essential and effective means of scientific development. It is only to emphasize the limitations, which follows when formal generalization becomes the only legitimate method of scientific inquiry.

The balanced view of the role of the case study in attempting to generalize by testing hypotheses has been formulated by Eckstein:

> *[C]omparative and case studies are alternative means to the end of testing theories, choices between which must be largely governed by arbitrary or practical, rather than logical, considerations . . .* [I]t is impossible to take seriously the position that case study is suspect because problem-prone and comparative study deserving of benefit of doubt because problem-free (1975, 116, 131, emphasis in original; see also Barzelay 1993, 305 ff.).

Eckstein here uses the term 'theory' in its 'hard' sense, that is, comprising explanation and prediction. This makes Eckstein's dismissal of the view that case studies cannot be used for testing theories or for generalization stronger than my own view, which is here restricted to the testing of 'theory' in the 'soft' sense, that is, testing propositions or hypotheses. Eckstein shows that if predictive theories would exist in social science, then the case study could be used to test these theories just as well as other methods.

More recently, John Walton (1992, 129) has similarly observed that 'case studies are likely to produce the best theory.' Eckstein observes, however, the striking lack of genuine theories within his own field, political science, but apparently fails to see why this is so:

> Aiming at the disciplined application of theories to cases forces one to state theories more rigorously than might otherwise be done--provided that the application is truly 'disciplined,' i.e., designed to show that valid theory compels a particular case interpretation and rules out others. As already stated, this, unfortunately, is rare (if it occurs at all) in political study. One reason is the lack of compelling theories (1975, 103-4).

The case study is ideal for generalizing using the type of test that Karl Popper called 'falsification,' which in social science forms part of critical reflexivity. Falsification is one of the most rigorous tests to which a scientific proposition can be subjected: if just one observation does not fit with the proposition it is considered not valid generally and must therefore be either revised or rejected. Popper himself used the now famous example of, 'All swans are white,' and proposed that just one observation of a single black swan would falsify this proposition and in this way have general significance and stimulate further investigations and theory-building. The case study is well suited for identifying 'black swans' because of its in-depth approach: what appears to be 'white' often turns out on closer examination to be 'black.'

Finding black swans was an experience with which I became thoroughly familiar when I did my first in-depth case study, of urban politics and planning in the city of Aalborg, Denmark (Flyvbjerg 1998a). For instance, in university I had been trained in the neoclassical model of 'economic man,' competition, and free markets. As I dug into what happened behind closed doors in Aalborg, I found that economic man does not live here. The local business community were power mongers who were busy negotiating illicit deals with politicians and administrators on how to block competition and the free market and create special privileges for themselves. The neoclassical model was effectively falsified by what I saw in Aalborg. Similarly, the model of representative democracy, which on the surface of things appears to apply, and by law is supposed to apply in Aalborg and Denmark, was strangely absent in the deep detail of the case. Here I found a highly undemocratic, semi-institutionalized way of making decisions, where leaders of the business community and of the city government had formed a secret council, which effectively replaced the democratically elected city council as the place where important decisions on urban politics and planning were made. My colleagues in third-world nations, who appear to hold less illusions about markets and democracy than academics in the first world, get a good laugh when I tell my Aalborg stories. They see that, after all, we in the North are not so different; we are third-world, too.

We will return to falsification in discussing the fourth misunderstanding of the case study below. For the present, however, we can correct the second misunderstanding--that one cannot generalize on the basis of a single case and that the case study cannot contribute to scientific development--so that it now reads:

> *One can often generalize on the basis of a single case, and the case study may be central to scientific development via generalization as supplement or alternative to other methods. But formal generalization is overvalued as a source of scientific development, whereas 'the force of example' is underestimated.*

**Strategies for Case Selection**

The third misunderstanding about the case study is that the case method is claimed to be most useful for generating hypotheses in the first steps of a total research process, while hypothesis-testing and theory-building is best carried out by other methods later in the process. This misunderstanding derives from the previous misunderstanding that one cannot generalize on the basis of individual cases. And since this misunderstanding has been revised as above, we can now correct the third misunderstanding as follows:

> *The case study is useful for both generating and testing of hypotheses but is not limited to these research activities alone.*

Eckstein--contravening the conventional wisdom in this area--goes so far as to argue that case studies are better for testing hypotheses than for producing them. Case studies, Eckstein (1975, 80) asserts, 'are valuable at all stages of the theory-building process, but most valuable at that stage of theory-building where least value is generally attached to them: the stage at which candidate theories are tested.' Testing of hypotheses relates directly to the question of 'generalizability', and this in turn relates to the question of case selection.

Here generalizability of case studies can be increased by the strategic selection of cases (on the selection of cases, further see Ragin 1992; Rosch 1978). When the objective is to achieve the greatest possible amount of information on a given problem or phenomenon, a representative case or a random sample may not be the most appropriate strategy. This is because the typical or average case is often not the richest in information. Atypical or extreme cases often reveal more information because they activate more actors and more basic mechanisms in the situation studied. In addition, from both an understanding-oriented and an action-oriented perspective, it is often more important to clarify the deeper causes behind a given problem and its consequences than to describe the symptoms of the problem and how frequently they occur. Random samples emphasizing representativeness will seldom be able to produce this kind of insight; it is more appropriate to select some few cases chosen for their validity.

Table A summarizes various forms of sampling. The *extreme case* can be well-suited for getting a point across in an especially dramatic way, which often occurs for well-known case studies such as Freud's 'Wolf-Man' and Foucault's 'Panopticon.' In contrast, a *critical case* can be defined as having strategic importance in relation to the general problem. For example, an occupational medicine clinic wanted to investigate whether people working with organic solvents suffered brain damage. Instead of choosing a representative sample among all those enterprises in the clinic's area that used organic solvents, the clinic strategically located a single workplace where all safety regulations on cleanliness, air quality, and the like, had been fulfilled. This model enterprise became a critical case: if brain damage related to organic solvents could be found at this particular facility, then it was likely that the same problem would exist at other enterprises which were less careful with safety regulations for organic solvents. Via this type of strategic choice, one can save both time and money in researching a given problem. Another example of critical case selection is the above-mentioned strategic selection of lead and feather for the test of whether different objects fall with equal velocity. The selection of materials provided the possibility to formulate a generalization characteristic of critical cases, a generalization of the sort, 'If it is valid for this case, it is valid for all (or many) cases.' In its negative form, the generalization would be, 'If it is not valid for this case, then it is not valid for any (or only few) cases.'

How does one identify critical cases? This question is more difficult to answer than the question of what constitutes a critical case. Locating a critical case requires experience, and no universal methodological principles exist by which one can with certainty identify a critical case. The only general advice that can be given is that when looking for critical cases, it is a good idea to look for either 'most likely' or 'least likely' cases, that is, cases which are likely to either clearly confirm or irrefutably falsify propositions and hypotheses. This is what I thought I was doing when planning the Aalborg case study mentioned above (Flyvbjerg 1998a). I was mistaken, however, but to my chagrin I did not realize this until I was halfway through the research process. Initially, I conceived of Aalborg as a 'most likely' critical case in the following manner: if rationality and urban planning were weak in the face of power in Aalborg then, most likely, they would be weak anywhere, at least in Denmark, because in Aalborg the rational paradigm of planning stood stronger than anywhere else. Eventually, I realized that this logic was flawed, because my research of local relations of power showed that one of the most influential 'faces of power' in Aalborg, the Chamber of Industry and Commerce, was substantially stronger than their equivalents elsewhere. This had not been clear at the outset because much less research existed on local power relations than research on local planning. Therefore, instead of a critical case, unwittingly I ended up with an extreme case in the sense that both rationality and power were unusually strong in Aalborg, and my case study became a study of what happens when strong rationality meets strong power in the arena of urban politics and planning. But this selection of Aalborg as an extreme case happened to me, I did not deliberately choose it. It was a frustrating experience when it happened, especially during those several months from when I realized I did not have a critical case until it became clear that all was not lost because I had something else. As a case researcher charting new terrain one must be prepared for such incidents, I believe.

A model example of a 'least likely' case is Robert Michels's (1962) classical study of oligarchy in organizations. By choosing a horizontally structured grassroots organization with strong democratic ideals--that is, a type of organization with an especially low probability of being oligarchical--Michels could test the universality of the oligarchy thesis; that is, 'If this organization is oligarchic, so are most others.' A corresponding model example of a 'most likely' case is W. F. Whyte's (1943) study of a Boston slum neighborhood, which according to existing theory should have exhibited social disorganization, but in fact showed quite the opposite (see also the articles on Whyte's study in *Journal of Contemporary Ethnography*, vol. 21, no. 1, 1992).

Cases of the 'most likely' type are especially well suited to falsification of propositions, while 'least likely' cases are most appropriate to tests of verification. It should be remarked that a most likely case for one proposition is the least likely for its negation. For example, Whyte's slum neighborhood could be seen as a least likely case for a hypothesis concerning the universality of social organization. Hence, the identification of a case as most or least likely is linked to the design of the study, as well as to the specific properties of the actual case.

A final strategy for the selection of cases is choice of the *paradigmatic case*. Thomas Kuhn has shown that the basic skills, or background practices, of natural scientists are organized in terms of 'exemplars' the role of which can be studied by historians of science. Similarly, scholars like Clifford Geertz and Michel Foucault have often organized their research around specific cultural paradigms: a paradigm for Geertz lay for instance in the 'deep play' of the Balinese cockfight, while for Foucault,

European prisons and the 'Panopticon' are examples. Both instances are examples of paradigmatic cases, that is, cases that highlight more general characteristics of the societies in question. Kuhn has shown that scientific paradigms cannot be expressed as rules or theories. There exists no predictive theory for how predictive theory comes about. A scientific activity is acknowledged or rejected as good science by how close it is to one or more exemplars; that is, practical prototypes of good scientific work. A paradigmatic case of how scientists do science is precisely such a prototype. It operates as a reference point and may function as a focus for the founding of schools of thought.

As with the critical case, we may ask, 'How does one identify a paradigmatic case?' How does one determine whether a given case has metaphorical and prototypical value? These questions are even more difficult to answer than for the critical case, precisely because the paradigmatic case transcends any sort of rule-based criteria. No standard exists for the paradigmatic case because it sets the standard. Hubert and Stuart Dreyfus see paradigmatic cases and case studies as central to human learning. In an interview with Hubert Dreyfus (author's files), I therefore asked what constitutes a paradigmatic case and how it can be identified. Dreyfus replied:

> Heidegger says, you recognize a paradigm case because it shines, but I'm afraid that is not much help. You just have to be intuitive. We all can tell what is a better or worse case--of a Cézanne painting, for instance. But I can't think there could be any rules for deciding what makes Cézannne a paradigmatic modern painter . . . [I]t is a big problem in a democratic society where people are supposed to justify what their intuitions are. In fact, nobody really can justify what their intuition is. So you have to make up reasons, but it won't be the real reasons.

One may agree with Dreyfus that intuition is central to identifying paradigmatic cases, but one may disagree it is a problem to have to justify one's intuitions. Ethnometholodogical studies of scientific practice have demonstrated that all variety of such practice relies on taken-for-granted procedures that feel largely intuitive. However, those intuitive decisions are accountable, in the sense of being sensible to other practitioners or often explicable if not immediately sensible. That would frequently seem to be the case with the selection of paradigmatic cases. We may select such cases on the basis of taken-for-granted, intuitive procedures but are often called upon to account for that selection. That account must be sensible to other members of the scholarly communities of which we are part. This may even be argued to be a general characteristic of scholarship, scientific or otherwise, and not unique to the selection of paradigmatic social scientific case studies. For instance, it is usually insufficient to justify an application for research funds by stating that one's intuition says that a particular research should be carried out. A research council ideally operates as society's test of whether the researcher can account, in collectively acceptable ways, for his or her intuitive choice, even though intuition may be the real, or most important, reason why the researcher wants to execute the project.

It is not possible consistently, or even frequently, to determine in advance whether or not a given case--Geertz' cock fights in Bali, for instance--is paradigmatic. Besides the strategic choice of case, the execution of the case study will certainly play a role, as will the reactions to the study by the research community, the group studied, and, possibly, a broader public. The value of the case study will depend on the validity claims which researchers can place on their study, and the status these claims obtain in dialogue with other validity claims in the discourse to which the study is a contribution. Like other good craftsmen, all that researchers can do is use their experience and intuition to assess whether they believe a given case is interesting in a paradigmatic context, and whether they can provide collectively acceptable reasons for the choice of case.

Finally, concerning considerations of strategy in the choice of cases, it should be mentioned that the various strategies of selection are not necessarily mutually exclusive. For example, a case can be simultaneously extreme, critical, and paradigmatic. The interpretation of such a case can provide a unique wealth of information, because one obtains various perspectives and conclusions on the case according to whether it is viewed and interpreted as one or another type of case.

**Do Case Studies Contain a Subjective Bias?**

The fourth of the five misunderstandings about case-study research is that the method maintains a bias toward verification, understood as a tendency to confirm the researcher's preconceived notions, so that the study therefore becomes of doubtful scientific value. Diamond (1996, 6), for example, holds this view. He observes that the case study suffers from what he calls a 'crippling drawback,' because it does not apply 'scientific methods,' by which Diamond understands methods useful for 'curbing one's tendencies to stamp one's pre-existing interpretations on data as they accumulate.'

Francis Bacon (1853, xlvi) saw this bias toward verification, not simply as a phenomenon related to the case study in particular, but as a fundamental human characteristic. Bacon expressed it like this:

> The human understanding from its peculiar nature, easily supposes a greater degree of order and equality in things than it really finds. When any proposition has been laid down, the human understanding forces everything else to add fresh support and confirmation. It is the peculiar and perpetual error of the human understanding to be more moved and excited by affirmatives than negatives.

Bacon certainly touches upon a fundamental problem here, a problem, which all researchers must deal with in some way. Charles Darwin (1958, 123), in his autobiography, describes the method he developed in order to avoid the bias toward verification:

> I had . . . during many years followed a golden rule, namely, that whenever a published fact, a new observation or thought came across me, which was opposed to my general results, to make a memorandum of it without fail and at once; for I had found by experience that such facts and thoughts were far more apt to escape from the memory than favorable ones. Owing to this habit, very few objections were raised against my views, which I had not at least noticed and attempted to answer.

The bias toward verification is general, but the alleged deficiency of the case study and other qualitative methods is that they ostensibly allow more room for the researcher's subjective and arbitrary judgment than other methods: they are often seen as less rigorous than are quantitative, hypothetico-deductive methods. Even if such criticism is useful, because it sensitizes us to an important issue, experienced case researchers cannot help but see the critique as demonstrating a lack of knowledge of what is involved in case-study research. Donald Campbell and others have shown that the critique is fallacious, because the case study has its own rigor, different to be sure, but no less strict than the rigor of quantitative methods. The advantage of the case study is that it can 'close in' on real-life situations and test views directly in relation to phenomena as they unfold in practice.

According to Campbell, Ragin, Geertz, Wieviorka, Flyvbjerg, and others, researchers who have conducted intensive, in-depth case studies typically report that their preconceived views, assumptions, concepts, and hypotheses were wrong and that the case material has compelled them to revise their hypotheses on essential points. The case study forces upon the researcher the type of falsifications described above. Ragin (1992, 225) calls this a 'special feature of small-$N$ research,' and goes on to explain that criticizing single-case studies for being inferior to multiple case studies is misguided, since even single-case studies 'are multiple in most research efforts because ideas and evidence may be linked in many different ways.'

Geertz (1995, 119) says about the fieldwork involved in most in-depth case studies that 'The Field' itself is a 'powerful disciplinary force: assertive, demanding, even coercive.' Like any such force, it can be underestimated, but it cannot be evaded. 'It is too insistent for that,' says Geertz. That he is speaking of a general phenomenon can be seen by simply examining case studies, like Eckstein (1975), Campbell (1975), and Wieviorka (1992) have done. Campbell (1975, 181-2) discusses the causes of this phenomenon in the following passage:

> In a case study done by an alert social scientist who has thorough local acquaintance, the theory he uses to explain the focal difference also generates prediction or expectations on dozens of other aspects of the culture, and he does not retain the theory unless most of these are also confirmed . . . Experiences of social scientists confirm this. Even in a single qualitative case study, the conscientious social scientist often finds no explanation that seems satisfactory. Such an outcome would be impossible if the caricature of the single case study . . . were correct--there would instead be a surfeit of subjectively compelling explanations.

According to the experiences cited above, it is falsification and not verification, which characterizes the case study. Moreover, the question of subjectivism and bias toward verification applies to all methods, not just to the case study and other qualitative methods. For example, the element of arbitrary subjectivism will be significant in the choice of categories and variables for a quantitative or structural investigation, such as a structured questionnaire to be used across a large sample of cases. And the probability is high that (1) this subjectivism survives without being thoroughly corrected during the study and (2) that it may affect the results, quite simply because the quantitative/structural researcher does not get as close to those under study as does the case-study researcher and therefore is less likely to be corrected by the study objects 'talking back.' According to Ragin:

> this feature explains why small-*N* qualitative research is most often at the forefront of theoretical development. When *N*'s are large, there are few opportunities for revising a casing [that is, the delimitation of a case]. At the start of the analysis, cases are decomposed into variables, and almost the entire dialogue of ideas and evidence occurs through variables. One implication of this discussion is that to the extent that large-*N* research can be sensitized to the diversity and potential heterogeneity of the cases included in an analysis, large-*N* research may play a more important part in the advancement of social science theory (1992, 225; see also Ragin 1987, 164-71).

Here, too, this difference between large samples and single cases can be understood in terms of the phenomenology for human learning discussed above. If one thus assumes that the goal of the researcher's work is to understand and learn about the phenomena being studied, then research is simply a form of learning. If one assumes that research, like other learning processes, can be described by the phenomenology for human learning, it then becomes clear that the most advanced form of understanding is achieved when researchers place themselves within the context being studied. Only in this way can researchers understand the viewpoints and the behavior, which characterizes social actors. Relevant to this point, Giddens states that valid descriptions of social activities presume that researchers possess those skills necessary to participate in the activities described:

> I have accepted that it is right to say that the condition of generating descriptions of social activity is being able in principle to participate in it. It involves 'mutual knowledge,' shared by observer and participants whose action constitutes and reconstitutes the social world (1982, 15).

From this point of view, the proximity to reality, which the case study entails, and the learning process which it generates for the researcher will often constitute a prerequisite for advanced understanding. In this context, one begins to understand Beveridge's conclusion that there are more discoveries stemming from the type of intense observation made possible by the case study than from statistics applied to large groups. With the point of departure in the learning process, we understand why the researcher who conducts a case study often ends up by casting off preconceived notions and theories. Such activity is quite simply a central element in learning and in the achievement of new insight. More simple forms of understanding must yield to more complex ones as one moves from beginner to expert.

On this basis, the fourth misunderstanding--that the case study supposedly contains a bias toward verification, understood as a tendency to confirm the researcher's preconceived ideas--is revised as follows:

> *The case study contains no greater bias toward verification of the researcher's preconceived notions than other methods of inquiry. On the contrary, experience indicates that the case study contains a greater bias toward falsification of preconceived notions than toward verification.*

**The Irreducible Quality of Good Case Narratives**

Case studies often contain a substantial element of narrative. Good narratives typically approach the complexities and contradictions of real life. Accordingly, such narratives may be difficult or impossible to summarize into neat scientific formulae, general propositions, and theories (Benhabib 1990, Rouse 1990, Roth 1989, White 1990, Mitchell and Charmaz 1996). This tends to be seen by critics of the case study as a drawback. To the case-study researcher, however, a particularly 'thick' and hard-to-summarize narrative is not a problem. Rather, it is often a sign that the study has uncovered a particularly rich problematic. The question, therefore, is whether the summarizing and generalization, which the critics see as an ideal, is always desirable. Nietzsche (1974, 335 [§ 373]) is clear in his answer to this question. 'Above all,' he says about doing science, 'one should not wish to divest existence of its *rich ambiguity*' (emphasis in original).

In doing the Aalborg study, I tried to capture the rich ambiguity of politics and planning in a modern democracy. I did this by focusing in-depth on the particular events that made up the case and on the minutiae that made up the events. Working with minutiae is time consuming, and I must concede that during the several years when I was toiling in the archives, doing interviews, making observations, talking with my informants, writing, and getting feedback, a nagging question kept resurfacing in my mind. This is a question bound to haunt many carrying out in-depth, dense case studies: 'Who will want to learn about a case like this, and in this kind of *detail*?' I wanted the Aalborg case study to be particularly dense because I wished to test the thesis that the most interesting phenomena in politics and planning, and those of most general import, would be found in the most minute and most concrete of details. Or to put the matter differently, I wanted to see whether the dualisms general-specific and abstract-concrete would metamorphose and vanish if I went into sufficiently deep detail. Richard Rorty has perceptively observed that the way to re-enchant the world is to stick to the concrete. Nietzsche similarly advocates a focus on 'little things.' Both Rorty and Nietzsche seem right to me. I saw the Aalborg case as being made up of the type of concrete, little things they talk about. Indeed, I saw the case itself as such a thing, what Nietzsche calls a discreet and apparently insignificant truth, which, when closely examined, would reveal itself to be pregnant with paradigms, metaphors, and general significance. That was my thesis, but theses can be wrong and case studies may fail. I was genuinely relieved when, eventually, the strategy of focusing on minutiae proved to be worth the effort.

Lisa Peattie (2001, 260) explicitly warns against summarizing dense case studies: 'It is simply that the very value of the case study, the contextual and interpenetrating nature of forces, is lost when one tries to sum up in large and mutually exclusive concepts.' The dense case study, according to Peattie is more useful for the practitioner and more interesting for social theory than either factual 'findings' or the high-level generalizations of theory.

The opposite of summing up and 'closing' a case study is to keep it open. Here I have found the following two strategies to work particularly well in ensuring such openness. First, when writing up a case study, I demur from the role of omniscient narrator and summarizer. Instead, I tell the story in its diversity, allowing the story to unfold from the many-sided, complex, and sometimes conflicting stories that the actors in the case have told me. Second, I avoid linking the case with the theories of any one academic specialization. Instead I relate the case to broader philosophical positions that cut across specializations. In this way I try to leave scope for readers of different backgrounds to make different interpretations and draw diverse conclusions regarding the question of what the case is a case of. The goal is not to make the case study be all things to all people. The goal is to allow the study to be different things to different people. I try to achieve this by describing the case with so many facets--like life itself--that different readers may be attracted, or repelled, by different things in the case. Readers are not pointed down any one theoretical path or given the impression that truth might lie at the end of such a path. Readers will have to discover their own path and truth inside the case. Thus, in addition to the interpretations of case actors and case narrators, readers are invited to decide the meaning of the case and to interrogate actors' and narrators' interpretations in order to answer that categorical question of any case study: 'What is this case a case of?'

Case stories written like this can neither be briefly recounted nor summarized in a few main results. The case story is itself the result. It is a 'virtual reality,' so to speak. For the reader willing to enter this reality and explore it inside and out the payback is meant to be a sensitivity to the issues at hand that cannot be obtained from theory. Students can safely be let loose in this kind of reality, which provides a useful training ground with insights into real-life practices that academic teaching often does not provide.

If we return briefly to the phenomenology for human learning we may understand why summarizing case studies is not always useful and may sometimes be counterproductive. Knowledge at the beginner's level consists precisely in the reduced formulas which characterize theories, while true expertise is based on intimate experience with thousands of individual cases and on the ability to discriminate between situations, with all their nuances of difference, without distilling them into formulas or standard cases. The problem is analogous to the inability of heuristic, computer-based expert systems to approach the level of virtuoso human experts, even when the systems are compared with the experts who have conceived the rules upon which these systems operate. This is because the experts do not use rules but operate on the basis of detailed case-experience. This is *real* expertise. The rules for expert systems are formulated only because the systems require it; rules are characteristic of expert *systems*, but not of real human *experts*.

In the same way, one might say that the rule formulation which takes place when researchers summarize their work into theories is characteristic of the culture of research, of researchers, and of theoretical activity, but such rules are not necessarily part of the studied reality constituted by Bourdieu's (1977, 8, 15) 'virtuoso social actors.' Something essential may be lost by this summarizing--namely the possibility to understand virtuoso social acting, which, as Bourdieu has shown, cannot be distilled into theoretical formulae--and it is precisely their fear of losing this 'something,' which makes case researchers cautious about summarizing their studies. Case researchers thus tend to be skeptical about erasing phenomenological detail in favor of conceptual closure.

Ludwig Wittgenstein shared this skepticism. According to Gasking and Jackson, Wittgenstein used the following metaphor when he described his use of the case study approach in philosophy:

> In teaching you philosophy I'm like a guide showing you how to find your way round London. I have to take you through the city from north to south, from east to west, from Euston to the embankment and from Piccadilly to the Marble Arch. After I have taken you many journeys through the city, in all sorts of directions, we shall have passed through any given street a number of times--each time traversing the street as part of a different journey. At the end of this you will know London; you will be able to find your way about like a born Londoner. Of course, a good guide will take you through the more important streets more often than he takes you down side streets; a bad guide will do the opposite. In philosophy I'm a rather bad guide (1967, 51).

This approach implies exploring phenomena firsthand instead of reading maps of them. Actual practices are studied before their rules, and one is not satisfied by learning only about those parts of practices that are open to public scrutiny; what Erving Goffman (1963) calls the 'backstage' of social phenomena must be investigated, too, like the side streets which Wittgenstein talks about.

With respect to intervention in social and political affairs, Abbott (1992, 79) has rightly observed that a social science expressed in terms of typical case narratives would provide 'far better access for policy intervention than the present social science of variables.' MacIntyre (1984, 216) similarly says, 'I can only answer the question 'What am I to do?' if I can answer the prior question 'Of what story or stories do I find myself a part?'' Several observers have noted that narrative is an ancient method and perhaps our most fundamental form for making sense of experience. (Novak 1975, 175; Mattingly 1991, 237; see also Abbott 1992, Arendt 1958, Carr 1986, Ricoeur 1984, Fehn et al. 1992, Rasmussen 1995, and Bal 1997).

To MacIntyre (1984, 214, 216), the human being is a 'story-telling animal,' and the notion of a history is as fundamental a notion as the notion of an action. In a similar vein, Mattingly (1991, 237) points out that narratives not only give meaningful form to experiences we have already lived through. They also provide us a forward glance, helping us to anticipate situations even before we encounter them, allowing us to envision alternative futures. Narrative inquiries do not--indeed, cannot--start from explicit theoretical assumptions. Instead, they begin with an interest in a particular phenomenon that is best understood narratively. Narrative inquiries then develop descriptions and interpretations of the phenomenon from the perspective of participants, researchers, and others.

Labov (1966, 37-9) writes that when a good narrative is over 'it should be unthinkable for a bystander to say, 'So what'?' Every good narrator is continually warding off this question. A narrative that lacks a moral that can be independently and briefly stated, is not necessarily pointless. And a narrative is not successful just because it allows a brief moral. A successful narrative does not allow the question to be raised at all. The narrative has already supplied the answer before the question is asked. The narrative itself is the answer (Nehamas 1985, 163-64).

A reformulation of the fifth misunderstanding, which states that it is often difficult to summarize specific case studies into general propositions and theories, thus reads as follows:

> *It is correct that summarizing case studies is often difficult, especially as concerns case process. It is less correct as regards case outcomes. The problems in summarizing case studies, however, are due more often to the properties of the reality studied than to the case study as a research method. Often it is not desirable to summarize and generalize case studies. Good studies should be read as narratives in their entirety.*

It must again be emphasized that despite the difficulty or undesirability in summarizing case studies, the case study method in general can certainly contribute to the cumulative development of knowledge; for example, in using the principles to test propositions described above under the second and third misunderstandings.

**Conclusions**

Today, when students and colleagues present me with the conventional wisdom about case-study research--for instance that one cannot generalize on the basis of a single case or that case studies are arbitrary and subjective--I know what to answer. By and large, the conventional wisdom is wrong or misleading. For the reasons given above, the case study is a necessary and sufficient method for certain important research tasks in the social sciences, and it is a method that holds up well when compared to other methods in the gamut of social science research methodology.

When students ask me for reference to a good book on how to carry out case study research in practice, I usually recommend Robert Stake's (1995) *The Art of Case Study Research*. If the students are intellectually curious, I suggest they also read Charles Ragin and Howard Becker's (1992) *What is a Case?* Both books are first-rate and fit well with the views presented in this article.

Let me reiterate, however, that the revision of the five misunderstandings about case study research described above, should not be interpreted as a rejection of research which focuses on large random samples or entire populations; for example, questionnaire surveys with related quantitative analysis. This type of research is also essential for the development of social science; for example, in understanding the degree to which certain phenomena are present in a given group or how they vary across cases. The advantage of large samples is breadth, while their problem is one of depth. For the case study, the situation is the reverse. Both approaches are necessary for a sound development of social science.

Here as elsewhere, the sharp separation often seen in the literature between qualitative and quantitative methods is a spurious one. The separation is an unfortunate artifact of power relations and time constraints in graduate training; it is not a logical consequence of what graduates and scholars need to know to do their studies and do them well. In my interpretation, good social science is opposed to an either/or and stands for a both/and on the question of qualitative versus quantitative methods. Good social science is problem-driven and not methodology-driven, in the sense that it employs those methods which for a given problematic best help answer the research questions at hand. More often than not, a combination of qualitative and quantitative methods will do the task best. Fortunately, there seems currently to be a general relaxation in the old and unproductive separation of qualitative and quantitative methods.

This being said, it should nevertheless be added that the balance between case studies and large samples is currently biased in favor of the latter in social science, so biased that it puts case studies at a disadvantage within most disciplines. In this connection, it is worth repeating the insight of Thomas Kuhn that a discipline without a large number of thoroughly executed case studies is a discipline without systematic production of exemplars, and that a discipline without exemplars is an ineffective one. In social science more good case studies could help remedy this situation.

*Table A: Strategies for the Selection of Samples and Cases*

| Type of Selection | Purpose |
|---|---|
| A. Random selection | To avoid systematic biases in the sample. The sample's size is decisive for generalization. |
|     1. Random sample | To achieve a representative sample which allows for generalization for the entire population. |
|     2. Stratified sample | To generalize for specially selected sub-groups within the population. |
| B. Information-oriented selection | To maximize the utility of information from small samples and single cases. Cases are selected on the basis of expectations about their information content. |
|     1. Extreme/deviant cases | To obtain information on unusual cases, which can be especially problematic or especially good in a more closely defined sense. |
|     2. Maximum variation cases | To obtain information about the significance of various circumstances for case process and outcome; e.g., three to four cases which are very different on one dimension: size, form of organization, location, budget, etc. |
|     3. Critical cases | To achieve information which permits logical deductions of the type, 'if this is (not) valid for this case, then it applies to all (no) cases.' |
|     4. Paradigmatic cases | To develop a metaphor or establish a school for the domain which the case concerns. |

**Notes**

---

[i]. The quote is from the original first edition of the dictionary (1984). In the third edition (1994), a second paragraph has been added about the case study. The entry is still highly unbalanced, however, and still promotes the mistaken view that the case study is hardly a methodology in its own right, but is best seen as subordinate to investigations of larger samples.